\documentclass[%
reprint,
superscriptaddress,
]{revtex4-2}
\usepackage{dblfloatfix}    
\usepackage{chemformula}

\usepackage{graphicx}
\usepackage{dcolumn}
\usepackage{bm}
\usepackage{amsmath}
\usepackage{amssymb}
\usepackage{accents}
\graphicspath{{Figures/}}
\usepackage[colorlinks]{hyperref}
\usepackage{times}
\usepackage{color}
\usepackage{float}

\usepackage{placeins}  

\definecolor{orange}{rgb}{1,0.5,0}

\usepackage{siunitx}
\usepackage{amsmath}

\definecolor{fxwarning}{rgb}{0.8,0,0}
\usepackage[normalem]{ulem}  

\newcommand*\around{{\raise.17ex\hbox{$\scriptstyle\mathtt{\sim}$}}}

\makeatletter
\providecommand*{\diff}%
  {\@ifnextchar^{\DIfF}{\DIfF^{}}}
\def\DIfF^#1{%
  \mathop{\mathrm{\mathstrut d}}%
    \nolimits^{#1}\gobblespace}
\def\gobblespace{%
    \futurelet\diffarg\opspace}
\def\opspace{%
  \let\DiffSpace\!%
  \ifx\diffarg(%
    \let\DiffSpace\relax
  \else
    \ifx\diffarg[%
    \let\DiffSpace\relax
    \else
      \ifx\diffarg\{%
    \let\DiffSpace\relax
      \fi\fi\fi\DiffSpace}
\providecommand*{\Diff}%
  {\@ifnextchar^{\DDIfF}{\DDIfF^{}}}
\def\DDIfF^#1{%
  \mathop{\mathrm{\mathstrut D}}%
    \nolimits^{#1}\gobblespace}
\makeatother

\usepackage[capitalize]{cleveref}  
\begin{document}


\preprint{APS/123-QED}

\title{Photonic molecule based on coupled ring quantum cascade lasers}




\author{Sara Kacmoli}
\email{skacmoli@princeton.edu}
\affiliation{Department of Electrical and Computer Engineering, Princeton University, Princeton, NJ 08544 USA}
\author{Deborah L. Sivco}
\affiliation{Department of Electrical and Computer Engineering, Princeton University, Princeton, NJ 08544 USA}
\affiliation{Current address: Trumpf Photonics, Inc., 2601 US Highway 130, Cranbury, NJ 08512 USA}
\author{Claire F. Gmachl}
\affiliation{Department of Electrical and Computer Engineering, Princeton University, Princeton, NJ 08544 USA}


\begin{abstract}
 Photonic molecules – particular systems composed of coupled optical resonators – emulate the behavior of complex physical systems exhibiting discrete energy levels. In this work, we present a novel photonic molecule composed of two strongly coupled, mid-infrared ring quantum cascade lasers. We explore both experimentally and numerically the key features of the photonic molecule such as the energy level splitting of bonding and antibonding supermodes. Due to the large size of the resonators, the energy splitting results in bands containing tens of modes. Each of these modes is furthermore doubly degenerate with respect to the direction of propagation, namely clockwise and counterclockwise. We explore several methods to carefully break these symmetries of the system in a controlled manner by introducing spatial and temporal asymmetries in the pumping scheme of the ring lasers. By employing these techniques, we achieve a high degree of precision in the dynamic control of the photonic molecule. Owing to their inherent suitability for on-chip integration, this new class of devices may enable applications as varied as novel mid-infrared sensors or a rich playground for studying non-Hermitian photonics and quantum optics with quantum cascade lasers. 

\end{abstract}

                              
\maketitle

\section{Introduction}

Discrete energy level schemes are present in many physical systems. Photonic analogues of such systems can be constructed by coupling resonators together to form so-called \textit{photonic molecules}.
This nomenclature reflects the fact that these artificial structures mimic bonded atoms, resulting in hybridized energy levels that resemble those observed in molecules. Wavelength-scale photonic molecules have already enabled the investigation and discovery of a number of complex phenomena in quantum optics~\cite{hartmannStronglyInteractingPolaritons2006, bambaOriginStrongPhoton2011}, non-Hermitian photonics~\cite{doi:10.1126/science.1258480, brandstetterReversingPumpDependence2014} or condensed matter physics.~\cite{doi:10.1126/science.aar4005} They have been used to demonstrate new optical computing paradigms~\cite{doi:10.1126/science.aah5178} and sensing capabilities.~\cite{hodaeiEnhancedSensitivityHigherorder2017,boriskinaSelfreferencedPhotonicMolecule2010, zhangSingleNanoparticleDetection2015}  
In all of the examples above, photonic molecules are based on microcavities, where the energy differences between the individual modes of each resonator are much larger than the splitting between the hybridized modes. Consequently, these systems usually contain only \textit{two} discrete energy levels. 

In our work, we demonstrate a photonic molecule based on two large scale and strongly coupled ring quantum cascade (QC) lasers. Due to the size of the system, the free spectral range (FSR) of our resonators is small compared to the splitting between the energy levels. As a result, our photonic molecule exhibits a new regime of coupling, where two degenerate combs give rise to two discrete energy \textit{bands} instead of only two individual levels. In condensed matter, this is akin to spin splitting of d-level orbitals which occurs, for example, in transition metal dichalcogenides.~\cite{borgelTransitionMetalDOrbital2016,kosmiderLargeSpinSplitting2013, zhuGiantSpinorbitinducedSpin2011a} Each of the energy levels in our combs is also doubly-degenerate due to the rotational symmetry of the ring laser (namely, the clockwise (CW) and counterclockwise (CCW) propagation directions.) 
Morever, photonic molecules based on mid-infrared QC lasers have not yet been demonstrated. In the \textit{terahertz} QC platform two studies have explored coupled microdisks, where the resonator size is comparable to the wavelength, in order to study non-Hermitian photonics, demonstrating the reversal of pump dependence at an exceptional point.~\cite{faschingElectricallyControllablePhotonic2009a, brandstetterReversingPumpDependence2014} However, coupling for terahertz QC lasers is significantly easier than in the mid-infrared due to the long wavelengths and the often-used metal-metal waveguides of the former. Mid-infrared QC lasers are well suited to sensing and spectroscopy applications due to the ro-vibrational resonances that many molecules have in this wavelength range. They are also unique among semiconductor lasers with a fast gain recovery time of \around~ps scale. Partially owing to this fast gain recovery, QC lasers are also sources of electrically-pumped frequency combs.~\cite{hugiMidinfraredFrequencyComb2012a} Realizing photonic molecules with mid-infrared QC lasers thus introduces a new class of lasers which would enable the development of novel and enhanced sensors. It also presents an opportunity to leverage the unique attributes of QC lasers in studying non-Hermitian photonics and quantum optics.

In our recent work we have demonstrated efficient evanescent, waveguide-mediated, in-plane outcoupling from a mid-infrared ring QC laser. Due to their active nature, these couplers have interesting properties that enable control over the emission of the ring laser.~\cite{kacmoliUnidirectionalModeSelection2022} This development holds significance especially given the recent demonstrations of ring QC lasers as sources of frequency combs and solitons.~\cite{piccardoFrequencyCombsInduced2020a, mengMidinfraredFrequencyComb2020, mengDissipativeKerrSolitons2022} However, in-plane coupling is not limited to simply coupling light out from ring lasers, but it can also be utilized to couple multiple identical lasers together, which is the core of the here discussed work.

We make use of these active couplers to realize our photonic molecule system operating at a wavelength $\lambda~\around$~\SI{8}{\um} with efficient active outcoupling waveguides. We experimentally and theoretically demonstrate key characteristics of a photonic molecule such as splitting of the energy bands. We show that each discrete band corresponds to a bonding and antibonding supermode. These supermodes are the product of the lifting of the degeneracy of states caused by specific symmetries in the photonic molecule.
We explore the conditions under which these states are degenerate as well as controlled degeneracy and symmetry breaking mechanisms. The photonic molecule exhibits great stability against perturbations of the symmetry of the rotational modes and the supermodes of the system.

\section{Device design}

\begin{figure}[b]
\centering
\includegraphics[scale = 0.41]{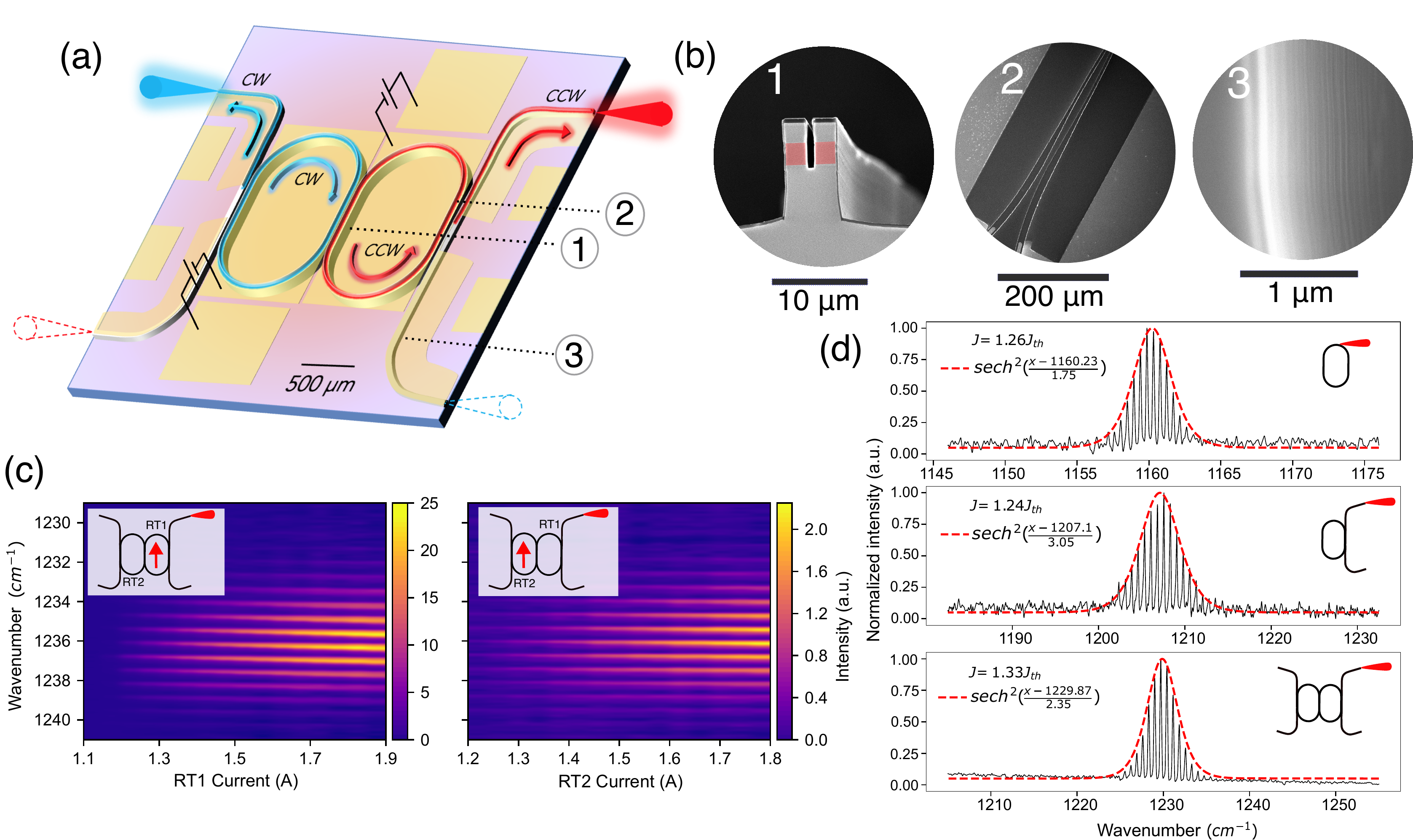}
\caption{(a) Diagram of the photonic molecule circuit, illustrating a representative steady-state supermode. Gold pads represent electrodes with top and ground contacts for each element (two pairs per outcoupling waveguide, one pair per racetrack).
    (b) SEM images detailing 1. a cross-section of the evanescent coupler between racetracks (false color marks the active region); 2. a top view of the taper geometry of the couplers, and 3. low sidewall roughness.
    (c) Measurements of lasing spectra at top-right port as a function of pumping current. Note that both racetracks exhibit nearly identical spectra over the entire current range. In the right plot, the right-hand racetrack is unpumped and serves as a passive waveguide coupling the optical output of the left-hand racetrack to the output port.
    (d) Measured spectra from three different devices: a standalone racetrack, a racetrack coupled to a waveguide, and a photonic molecule system as shown in (a). Note that the comb structure remains intact in all three configurations. The center wavelengths and free spectral range (FSR) values are different because these are devices of different sizes fabricated on different wafers.
}
\label{fig1}
\end{figure}

The photonic molecule system consists of two nominally identical racetrack lasers evanescently coupled by an air gap, each of which is in turn evanescently coupled to an active waveguide for the purpose of light extraction (Fig.~\ref{fig1}a). The design of the coupler sections is described in detail in our previous work~\cite{kacmoliUnidirectionalModeSelection2022}; it features adiabatic tapers which enhance the coupling strength  resulting in reasonable coupler lengths for air gaps of \around\SI{1}{\um} that are easily achievable by standard photolithography techniques. Scanning electron microscope (SEM) images of the air gap, the adiabatic taper and smooth sidewalls are presented in Fig.~\ref{fig1}{b}. The length of each racetrack is \SI{3.92}{\mm} and the coupler section between the two racetracks is \SI{190}{\um}.

Each racetrack can theoretically support independent counter-propagating modes. Due to the symmetry of the system, these modes are equally favored to achieve lasing. When the vertical sidewalls have minimal roughness, thus inducing minimal backscattering, the racetrack is unidirecional -- either lasing in the  clockwise (CW) or counter-clockwise (CCW) direction.
The two lasers exchange optical energy through the evanescent coupler between the CW (CCW) propagating mode of one ring and the CCW (CW) mode of the other. Therefore, thanks to unidirectionality, the rings strongly tend to oscillate in opposite orientations (e.g. as depicted in Fig.~\ref{fig1}a), resulting in two equally probable combinations: CW/CCW and CCW/CW. When measuring the light output at one of the ports in Fig.~\ref{fig1}a, only one of these combinations is in fact observed at a time, as we will show below.

The presence of a coupling mechanism itself can be interpreted as a source of potential loss and backscattering which raises a concern that adding these elements will perturb the internal ring dynamics responsible for frequency comb and dissipative Kerr soliton generation in recent work.~\cite{mengDissipativeKerrSolitons2022} To assess the degree of perturbation, we compare the spectra at similar pumping levels from three devices: a standalone racetrack laser, a racetrack with one waveguide coupler and a photonic molecule system all shown in Fig.~\ref{fig1}d. We find that they exhibit the same spectral envelope well captured by a characteristic $sech^2$ fit indicative of the presence of solitons. 

To experimentally assess the coupling efficiency we measure the output of each ring from the same waveguide arm in a device where the coupling between the rings and between each ring and the waveguides is nominally identical due to the same length and gap of the coupling region. In this measurement we keep all components of the system lightly pumped to overcome absorption except for the ring of interest, which is pumped above threshold. The results of this experiment are presented in Fig.~\ref{fig1}c. We note that the spectra of the two rings are remarkably similar. The estimated power coupling between the rings is $\around 8\%$. 

\onecolumngrid

\twocolumngrid
\null\cleardoublepage

\section{Photonic molecule ring QC lasers}

Having established important basic features of our ring lasers and couplers, we turn to the photonic molecule implementation of this system.
When two identical resonators with resonance frequency $f_0$ couple with strength $J$, it is well known that the resulting system exhibits two hybridized modes or \textit{supermodes} similarly to two atomic orbitals in a molecule. These two supermodes have two distinct frequencies (or energy levels), $f_0 \pm J$. The low energy mode is often called \textit{bonding}, whereas the high energy mode is called \textit{antibonding}.
Mathematically, this coupling can be modeled as a perturbation to the Hamiltonian operator of the system ($\mathcal{H}$), which causes a change in its eigenvalues (or, in this case, the frequency of the supermodes).

In an active system, gain and loss terms are also present in the Hamiltonian causing the eigenvalues to have positive or negative imaginary parts. Taking these effects into consideration, we use the following Hamiltonian matrix to describe the photonic molecule system, inspired by previous studies~\cite{kimDirectObservationExceptional2016}:

\begin{widetext}
\begin{equation}
    i\frac{\diff}{\diff t}
    \begin{pmatrix}
    E_1 \\
    E_2
    \end{pmatrix}
    =
    \mathcal{H}\begin{pmatrix}
    E_1 \\
    E_2
    \end{pmatrix}
    = 
    2\pi\begin{pmatrix}
    f_1 + i(\gamma_1 - \kappa_1) & J \\
    J & f_2 + i(\gamma_2 - \kappa_2)
    \end{pmatrix}
    \begin{pmatrix}
    E_1 \\
    E_2
    \end{pmatrix}
    .
    \label{eq:hamiltonian}
\end{equation}
\end{widetext}

In Eq.~\ref{eq:hamiltonian}, $E_1$ and $E_2$ are complex numbers representing the amplitude of fields in each cavity, $f_1$ and $f_2$ are the center resonance frequencies of the two resonators, $\gamma_1, \kappa_1$ and $\gamma_2, \kappa_2$ represent the gain and loss respectively for each cavity. We have allowed for the possibility of gain and loss imbalance as well as non-identical central frequency which is natural, albeit small, in our system given fabrication variation in the geometry of the resonators.
The Hamiltonian admits two eigenvalues ($f_{\pm}$) and corresponding eigenvectors ($\lambda_\pm$) -- or ``eigenfrequencies'' and ``supermodes'', respectively -- as shown in Eqs.~\ref{eq:eigenvals}~and~\ref{eq:eigenvecs}:

\begin{align}
    f_{\pm} = &f_{0} + i\eta_{0} \pm \sqrt{\Delta f^2 + 2i\Delta f \Delta \eta + J^2 - \Delta \eta^2}\nonumber\\
    &\approx f_{0} + i\eta_{0} \pm \sqrt{J^2 - \Delta \eta^2}\textrm{,  if } \Delta f \ll J%
    \label{eq:eigenvals}
\end{align}

\begin{equation}
    \lambda_\pm
    \equiv
    \begin{pmatrix}
    E_1 \\
    E_2
    \end{pmatrix}_\pm 
    =
    \begin{pmatrix}
    (\Delta f + i \Delta\eta + f_\pm - f_0 - i\eta_0)/J \\
    1
    \end{pmatrix}%
    \label{eq:eigenvecs}
\end{equation}
where we have defined the following quantities for clarity:
\begin{align*}
    f_{0} = \frac{f_1+f_2}{2},\quad
    \eta_{0} = \frac{(\gamma_1-\kappa_1)+(\gamma_2-\kappa_2)}{2},\\
    \Delta f = \frac{f_1-f_2}{2},\quad
    \Delta \eta = \frac{(\gamma_1-\kappa_1)-(\gamma_2-\kappa_2)}{2}.
\end{align*}


In Eq.~\ref{eq:eigenvals}, $\Delta\eta$ represents the imbalance in the gain-loss contrast between the two rings, and $\Delta f$ the natural center resonance frequency difference between resonators. In the case of our photonic molecule, $\Delta f \ll J$.

\begin{figure}[ht]
\centering
\includegraphics[scale = 0.4]{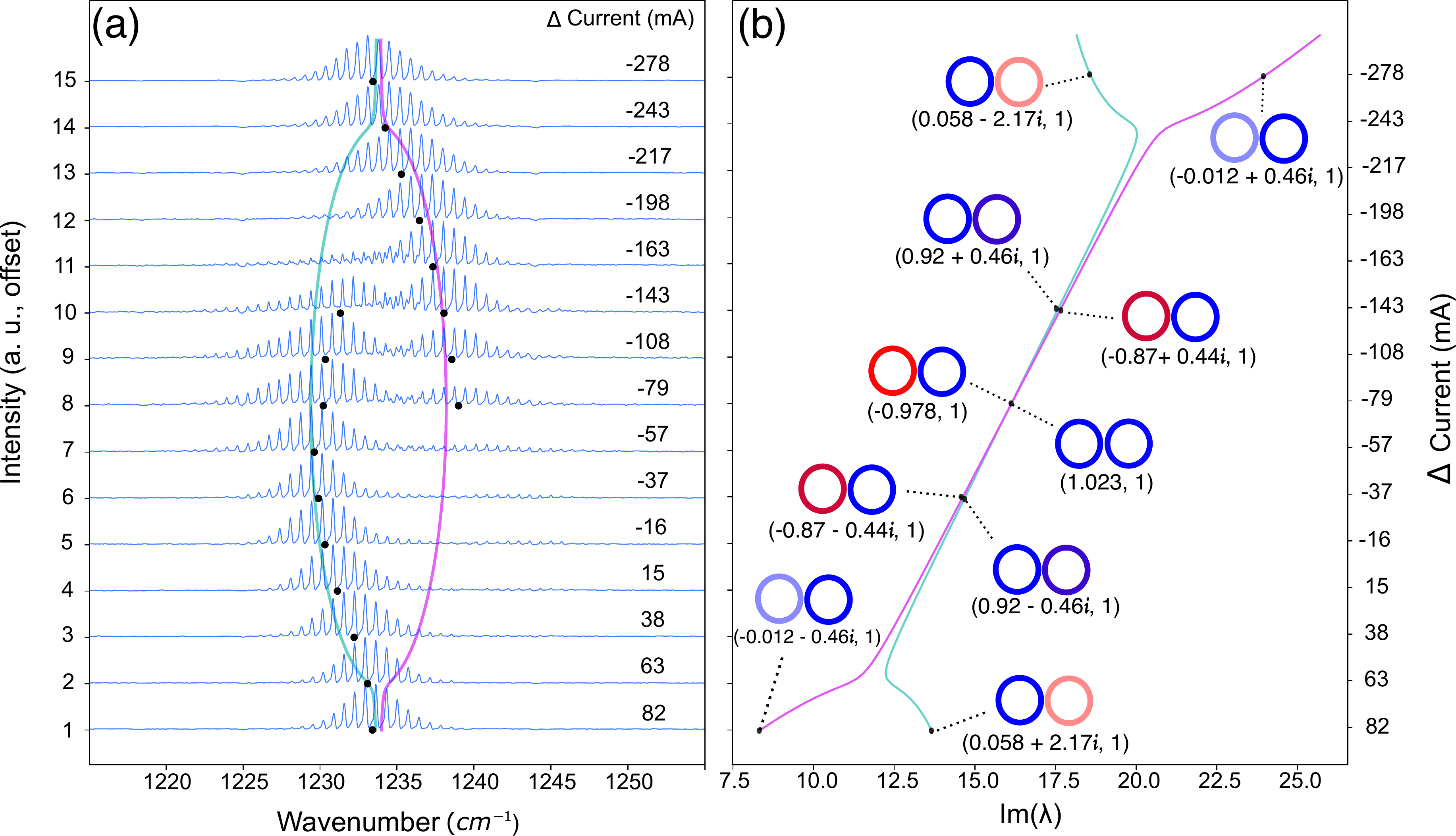}
\caption{(a) Spectral evolution of the photonic molecule as a function of current imbalance between the two racetracks (blue). Black dots represent the centers of the spectral envelopes. Cyan and magenta lines represent a fit of the real part of the two eigenfrequencies calculated based on Eq.~\ref{eq:eigenvals}. (b) Cyan and magenta lines represent the corresponding imaginary part of the two eigenfrequencies. For several different values of current imbalance we depict the eigenvectors symbolically where the color represents the sign of the eigenvector (blue, positive and red, negative) and the saturation represents its amplitude. The computed complex eigenvectors are given below each supermode. Additional simulation details including fitting parameters are given in the Supplementary Material.}
\label{fig:fig2}
\end{figure}

Experimentally, the ring lasers are pumped electrically, thus the symmetry/asymmetry of the pumping determines the modal gain terms ($\gamma_1$ and $\gamma_2$ modified by a confinement factor). Therefore, we expect $\Delta\eta$ to be linearly dependent on the current injection imbalance applied between the rings.
The two supermodes can be observed by measuring the optical spectrum emanating from the photonic molecule at any of its outcoupling ports as a function of pump current imbalance.
According to Eq.~\ref{eq:eigenvals}, we would expect a split between the frequencies of the supermode when $\Delta\eta < J$, and the collapse to a central frequency otherwise.
As shown in \cref{fig:fig2}{a}, we find a gradual splitting between the observed frequencies of the photonic molecule system. When the gain loss imbalance between the two racetracks ($\Delta\eta$) is zero, we expect the largest splitting between the two modes corresponding to be approximately $J$.
This situation occurs for a current imbalance of \\$\around -\SI{79}{\mA}$, a small value relative to an average current of \SI{1.45}{\A}, which we attribute to a slight loss term ($\kappa$) asymmetry between two ring lasers.

Based on Eq.~\ref{eq:eigenvals}, we fit the real and imaginary parts of the two calculated eigenfrequencies (\cref{fig:fig2}). The real part of the eigenfrequency shown in~\cref{fig:fig2}a explains the frequency split present in the photonic molecule spectrum. However, we note that for $\Delta I > -\SI{79}{\mA}$, the lowest frequency supermode dominates. This can be explained by the imaginary part of the eigenfrequency, which corresponds to the gain asymmetry between the two supermodes.
When the imaginary parts of the eigenfrequencies shown in ~\cref{fig:fig2}b are equal, the photonic molecule will evolve to either supermode with equal probability. However, when one eigenfrequency has larger imaginary part than the other, we expect that the photonic molecule is more likely to converge to that supermode instead. In this device, since $\Delta f \neq 0$, the imaginary parts of the eigenfrequency are not equal (Eq.~\ref{eq:eigenvals} and \cref{fig:fig2}{b}).
When $\Delta\eta$ is sufficiently large, the two curves diverge and one supermode clearly dominates over the other.
This divergence can be understood by examining the values of the eigenvectors (visualized symbolically in \cref{fig:fig2}{b}, where the color represents the sign of the eigenvector and the saturation represents the amplitude). The amplitude and phase of the eigenvectors for the entire range of $\Delta \eta$ are given in Figs.~S1 and S2.
In the lower-frequency supermode, the fields in both rings are in-phase; whereas the higher-frequency supermode, the fields have opposite phase. This is analogous to symmetric and anti-symmetric molecular states that exist in atomic molecular bonds, where the symmetric state, called `bonding', has lower energy. Although many systems that exhibit these two states show a preference for the lower energy one, both states are observed in our system. We later also explore mechanisms that allow us to select one of them.

\clearpage
\onecolumngrid

\twocolumngrid
To gain some insight into the physical origin of this energy split, we consider the intensity distribution of the two supermodes. Bonding supermodes have an increased intensity of the electric field in the air gap compared to antibonding supermodes which have a node there (see Fig.~S3). Thus, the effective refractive indices of these two modes are different, leading to the frequency shift between the two supermodes. We have calculated these two modes (shown in Fig. S3), and their effective refractive indices are different by 0.45$\%$ which corresponds to a \SI{5.54}{\cm^{-1}} shift in frequency. This is, in fact, a close match for our observed split of \around~\SI{8}{\cm^{-1}}.
When $\Delta\eta$ exceeds a bonding threshold $\around J$, i.e. $\Delta\eta > J$, the photonic molecule dissociates, and the eigenmodes converge to individual ring modes.

\section{Rotational Symmetry Breaking}

 The two supermodes discussed in the previous section fundamentally originate from a degree of freedom in the phase between the fields in each ring. The bonding and antibonding supermodes physically manifest as fields that constructively or destructively interfere at the coupling region, respectively.
Furthermore, the photonic molecule exhibits a degeneracy related to the propagation orientation of the fields (namely CW and CCW) in each ring. Due to the rotational symmetry of the ring, both orientations have the same $\gamma$ and $\kappa$ values.
Since we have established from experiment that each ring is unidirectional, there are theoretically four possible mutually exclusive propagation orientations: CW/CW, CCW/CCW, CW/CCW and CCW/CW. Accounting for this extra degree of freedom, the two fields $E_{1}$ and $E_{2}$ from the previous section can be further distinguished in four field quantities, $E_1^{CW}$, $E_1^{CCW}$, $E_2^{CW}$, $E_2^{CCW}$.
We can observe these four outputs experimentally by placing detectors at the end of each waveguide arm.
When accounting for the mirror symmetry in the coupled-ring geometry, these four orientations can be classified into two equivalent modes: CW/CW and CCW/CCW (namely `competing'); and CW/CCW and CCW/CW (namely `reinforcing'). Since the evanescent directional coupler transfers energy from the CW mode of one ring into the CCW mode of the other, and vice versa, we expect that when the photonic molecule oscillates in a CW/CCW (or CCW/CW) configuration, the fields `reinforce' each other. Otherwise, in the `competing' configuration (CW/CW or CCW/CCW), the coupler transfers energy from one ring into a counterpropagating mode of the other, which is then damped under the much stronger laser mode.
\begin{figure}[ht]
\centering
\includegraphics[scale = 0.51]{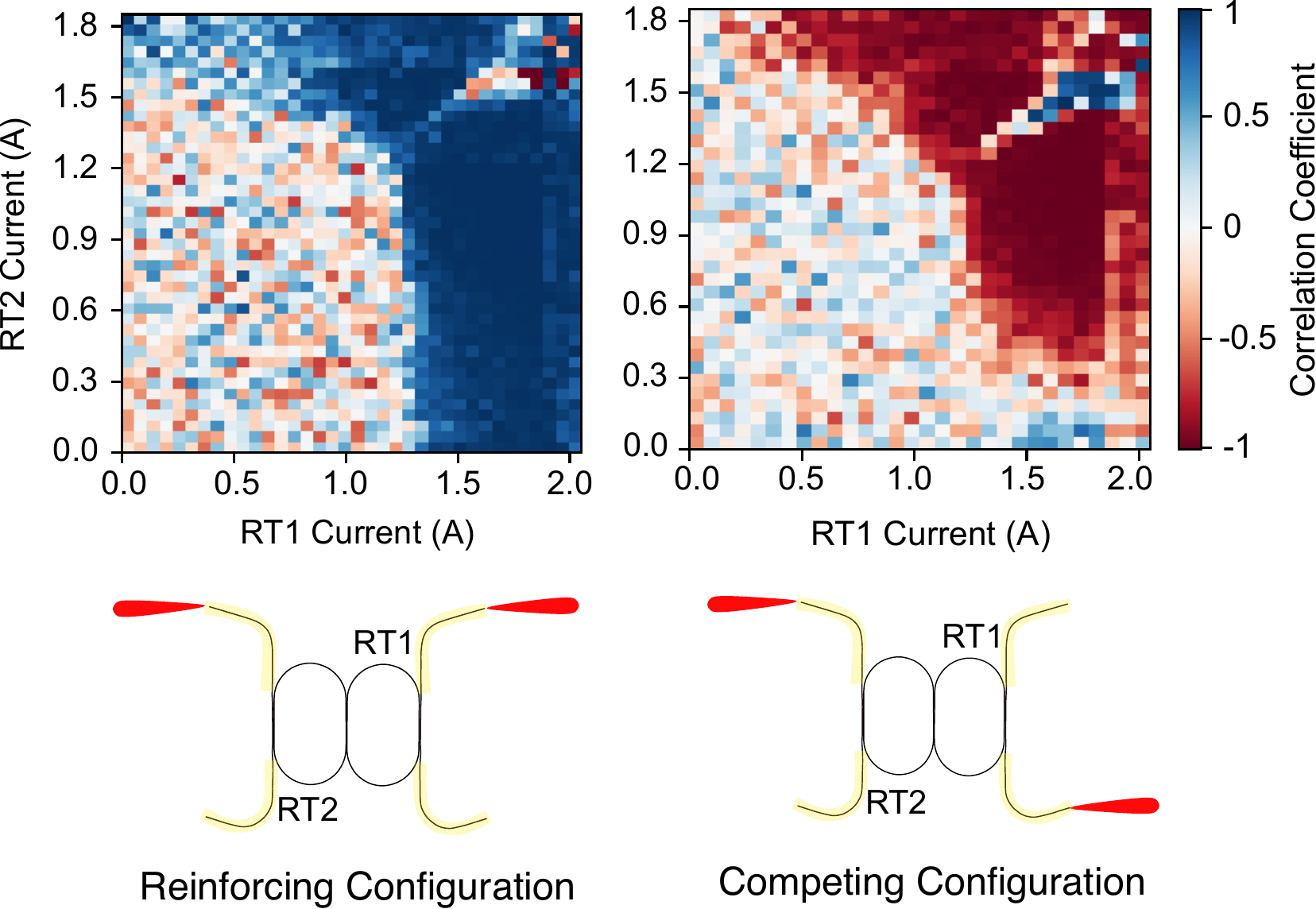}
\caption{Pearson's correlation coefficient of light pulses measured in the reinforcing configuration (left panel) and competing configuration (right panel) for a 2D sweep of currents applied to the two racetracks. Below the lasing threshold of each racetrack, the photodetector does not capture a signal, hence the correlation values are random. The schematics of each measurement configuration depicting the detector placement in red are shown below each panel. }
\label{fig:fig3new}
\end{figure}
To demonstrate this effect, we place one detector in the CW port of one ring and the CCW port of the other. When both rings are pumped above their lasing threshold at the same time, whenever an output pulse in one CW port is detected, another pulse is detected in the opposite CCW port. This is shown as a positive correlation in the left panel of \cref{fig:fig3new}. Conversely, when the detectors are placed in a CW/CW configuration, these two outputs are anticorrelated, presented in the right panel of \cref{fig:fig3new}.
As this figure shows, this behavior occurs regardless of the pump levels applied to the two racetracks, indicating that this behavior is inherent to the photonic molecule. One exception to this rule is when the pump currents are nearly identical. In this case we observe that the correlation disappears.
This is the same regime where we find the probabilistic switching between the two supermodes discussed in the previous section. Generally, we find that the split in the frequency, i.e. operating in the supermode regime, is accompanied by rapid fluctuations of the photodetector signal, leading to the distinct change in correlation along the diagonal. This is due to the rotational mode degeneracy: when the two rings are pumped equally, there is a probabilisitic choice between the CW/CCW and CCW/CW configurations of the supermodes.

\begin{figure}[ht]
\centering
\includegraphics[scale = 0.49]{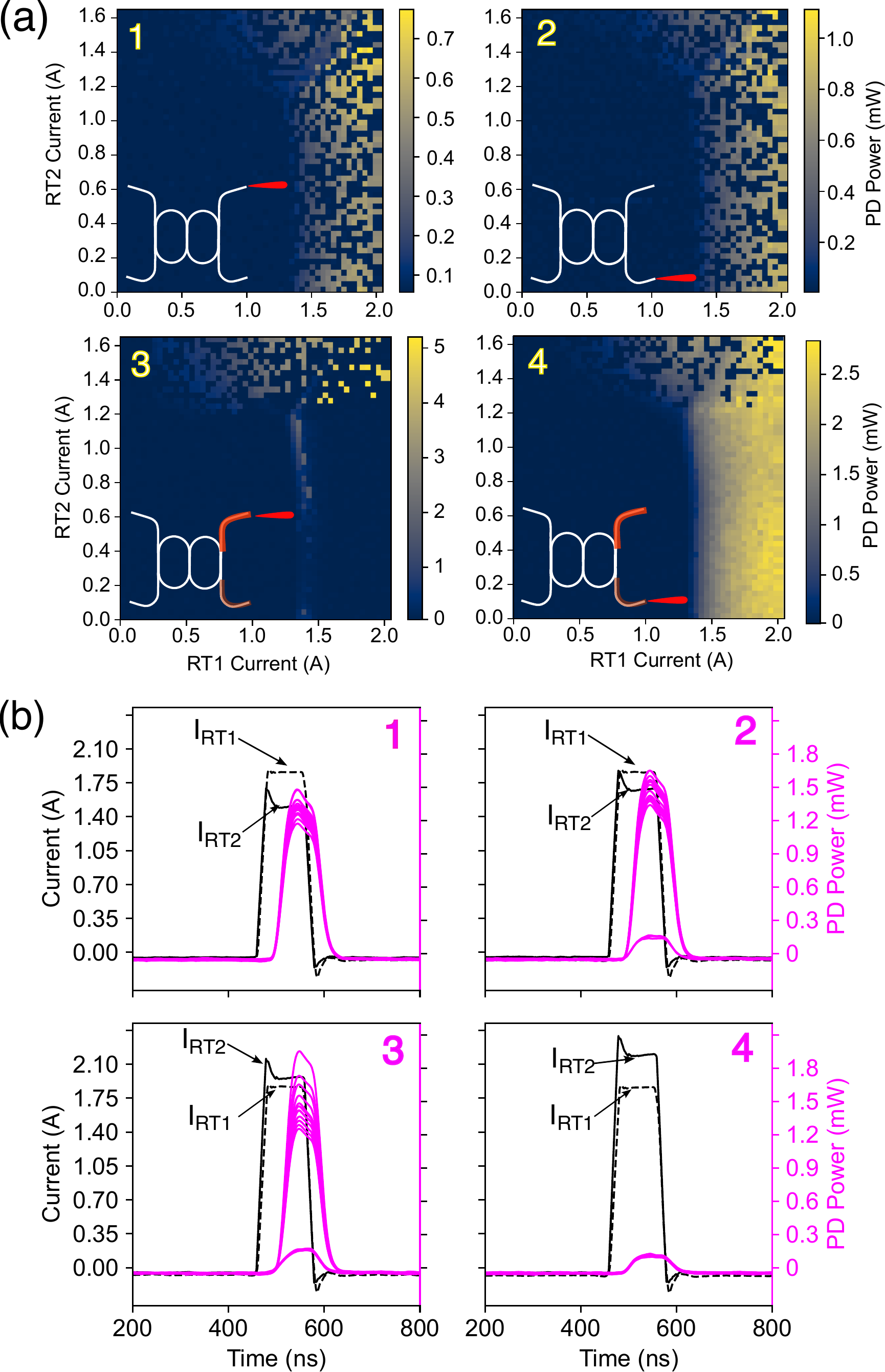}
\caption{(a) Panels 1 and 2 show the photodetector power measured at the CCW and CW ports of racetrack 1, respectively, when no mode selection mechanism is applied. Panels 3 and 4 show the photodetector power measured at the CCW and CW ports of racetrack 1, respectively, when a small current imbalance of \around \SI{75}{\mA} is applied to the waveguide arms thus favoring the CW mode. (b) Current pulses of the two racetracks (black solid and black dashed) along with 20 consecutive photodetector traces measured at the CW port of racetrack 1 (magenta). In panels 1 through 4 we gradually increase the applied pump power to racetrack 2 while keeping racetrack 1 at a constant pumping level. As the pumping level of racetrack 2 exceeds that of racetrack 1, the output switches from fully CW to fully CCW.}
\label{fig:fig4new}
\end{figure}

Because of the pulsed nature of our pumping scheme and the symmetry described above, each racetrack chooses one direction of emission randomly for every pulse. Therefore, a typical 2D sweep of the racetrack currents $I_\textrm{RT1}$, $I_\textrm{RT2}$ yields the results shown in \cref{fig:fig4new}{a1} and \cref{fig:fig4new}{a2}. The `pixelation' of the output power in these figures indicates this random selection. However, we can break this rotational symmetry by asymmetrically pumping the outcoupling waveguide arms as we have shown in detail in our previous work.~\cite{kacmoliUnidirectionalModeSelection2022} This allows us to deterministically select a particular direction of emission. For example, in \cref{fig:fig4new}{a3} and \cref{fig:fig4new}{a4}, we have used this mechanism to select the CW mode of racetrack 1 as demonstrated by the lack of signal in the CCW port (panel 3) and full signal in the CW port (panel 4).

With this mechanism, we can perturb the natural symmetry of our photonic molecule and observe the steady state intensity distribution that it selects. We configure the racetracks in a competing arrangement, such that both racetracks prefer to emit in CW. We show 20 consecutive photodetector traces measured on the CW port of racetrack 1 in \cref{fig:fig4new}{b} along with the pumping levels of the two racetracks. We note that when the current in racetrack 1 is higher than racetrack 2 (panel 1), the output pulses are always CW as we discussed above. As the current levels become more similar, we begin to see the signal fluctuate between the two ports (panels 2 and 3). Finally, when the current in racetrack 2 is higher than in racetrack 1, the output switches fully to CCW (panel 4). We conclude that the photonic molecule only oscillates in the `reinforcing' configuration, even in the presence of external stimulation. In this experiment, the photonic molecule never converges to a `competing' configuration.

\section{Stability of Photonic Molecule Supermodes}

\begin{figure}[ht]
\centering
\includegraphics[scale = 0.37]{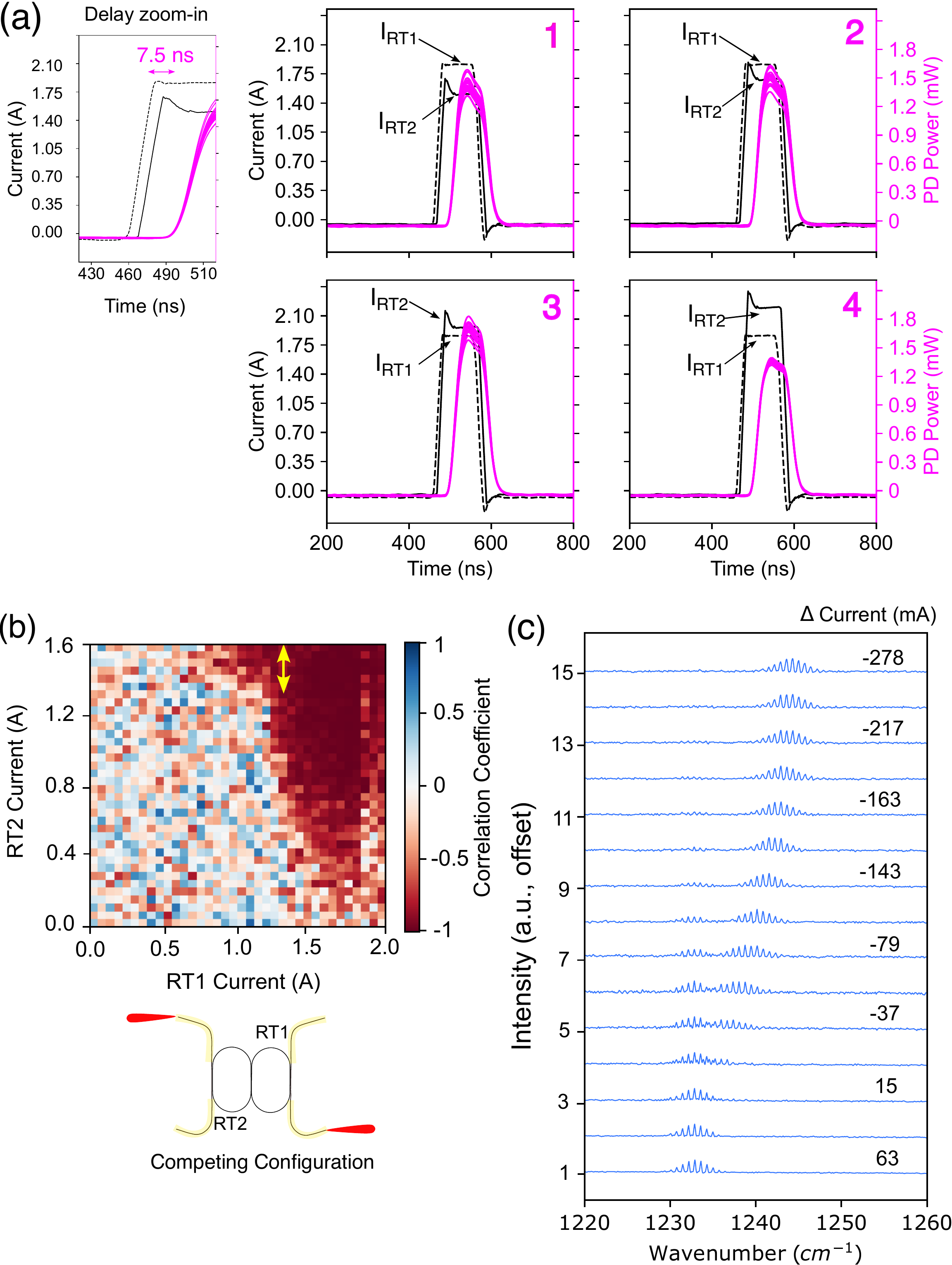}
\caption{(a) Current pulses of the two racetracks (black solid and black dashed) along with 20
consecutive photodetector traces measured at the CW port of racetrack 1 (magenta). As the pumping level of racetrack 2 exceeds that of racetrack 1 (shown in the transition from panel 1 to 4), the output remains CW due to the presence of a small delay between the pulses where racetrack 1 is ahead of racetrack 2 by \SI{7.5}{\ns}. A magnified view of the leading edge of the pulses to highlight the delay is shown on the left panel. (b) Pearson's correlation coefficient for a competing configuration in the case of a \SI{7.5}{\ns} delay. Note how the instability in the correlation around the diagonal, which we observe in~\cref{fig:fig3new}a,  disappears for this plot. The yellow arrow corresponds to the figure part (c). (c) Spectral evolution of the photonic molecule when a small delay is introduced between the two racetrack lasers. This measurment is taken for the region marked with a yellow arrow in (b) where we would expect to find correlation instability and a frequency split. Instead, due to the careful current pulse timing, the system stably selects one of the supermodes. }
\label{fig:fig5}
\end{figure}

In all previous experiments, both lasers were electrically pumped via synchronous pulses of widths varying from 50-100 ns with a repetition rate of \SI{60}{\kHz}. With every pulse, the rings transition from a zero-field state and converge to one of two possible supermodes ($\lambda_{+}$ or $\lambda_{-}$), as well as one of two allowed rotational configurations (CW/CCW or CCW/CW).
Here, we probe the temporal dynamics of our system thanks to the pulsed pumping scheme. We find that once the photonic molecule converges to a state, that state is robust to perturbations. To demonstrate this phenomenon, we set up a similar experiment as shown in the rotational symmetry breaking case (refer to \cref{fig:fig4new}b): one of the lasers (in this case racetrack 1) is set up to operate CW. Unlike the previous experiment; however, this laser's pump pulse leads the second racetrack's pump pulse by \SI{7.5}{\ns}.
During this period, the field in the racetrack 1 cavity starts oscillating only in the CW orientation as shown by the photodetector traces in \cref{fig:fig5}{a1}. 
Differently from what we described in the previous section (\cref{fig:fig4new}{b}), even as the pump amplitude of racetrack 2 (also primed for CW operation) becomes greater, racetrack 1 still chooses only the CW mode as shown by the photodetector traces in \cref{fig:fig5}{a4}, forcing racetrack 2 to operate CCW. The takeaway from this experiment is that the propagation orientation is determined at the onset of the current pulse and is stable against perturbations.

We have also found that a small delay induced in this way will remove any instabilities in the upper-right quadrant of the correlation plots which we previously observed in \cref{fig:fig3new} due to the stable rotational mode selection. One such correlation plot where the two racetracks have a small delay between them is shown in (\cref{fig:fig5}{b}). We note that along the diagonal, pulses show full anticorrelation due to the competing configuration setup. 
In addition to lifting this degeneracy between the rotational modes, we postulate that the delay also lifts the equiprobable fluctuation between the two supermodes. This is supported by measuring the spectra at one section of the diagonal (depicted with a yellow arrow in \cref{fig:fig5}b). We perform the same measurement as shown in \cref{fig:fig2}a, where the two ring lasers are pumped asymmetrically, but now with the addition of a small delay between the racetrack pulses. The result of this experiment is presented in \cref{fig:fig5}c. We find a lack of the probabilistic region where the system constantly switches between the two supermodes. Instead, the system converges from an individual ring mode to one of the two possible supermodes, revealing a one-sided spectrum (compare to the two-sided spectrum in \cref{fig:fig2}a).  
From this we can conclude that the photonic molecule can be primed in a particular initial condition to select a rotational state or a supermode in two ways: spatially, by judicious pumping imbalances of different elements as well as temporally, by introducing small delays between the two lasers.

\section{Conclusion}

In summary, we have demonstrated a mid-infrared photonic molecule based on coupled ring quantum cascade lasers. Evanescently-coupled active waveguides are used for in-plane light detection, extracting \around\SI{5}{\mW} of output power from these pulsed, room-temperature devices, without significantly perturbing the inner dynamics of the ring. We have measured the same spectral envelope and observed unidirectional lasing in several configurations of ring lasers –– standalone, coupled, and photonic molecule.
The photonic molecule shows spectral mode splitting in bonding and antibonding supermodes in good agreement with theoretical predictions. Due to the large size of our resonators, combs of frequencies are supported and instead of two individual levels, our photonic molecule exhibits a splitting of two energy bands each containing \around 10 modes. This is a novel regime of coupling for photonic molecules, where the FSR of the cavity is much smaller than the coupling strength. Additionally, each mode of the photonic molecule is also doubly degenerate due to the rotational symmetry in the propagation direction -- namely CW and CCW. Due to a combination of the control afforded by waveguide pumping and assymetric puming of the rings, we can spatially control the rotational symmetry breaking. Through a series of experiments where we control the selection of this rotational mode of each ring, we have shown that the photonic molecule always oscillates in a `reinforcing' configuration. By introducing a small temporal delay, we find that both the rotational mode and the supermode selection in the photonic molecule is determined right at the onset of pumping and is robust against perturbations. The high degree of dynamic control we have demonstrated in this work coupled with the unique properties of QC lasers renders these devices a testbed for exploration of non-Hermitian photonics, quantum optics and coherent instabilities in coupled laser systems. Our analysis of the photonic molecule provides insight on a new class of devices in the mid-infrared regime. Given their inherent compatibility with photonic integration, these devices pave the way towards compact, chip-scale mid-infrared sensors with enhanced sensitivity.

\newenvironment{backmatter}{%
  \small%
  \newcommand{\bmsection}[1]{\par\medskip\noindent{\bfseries ##1.\enspace}}%
}{}









\begin{thebibliography}{19}%
\makeatletter
\providecommand \@ifxundefined [1]{%
 \@ifx{#1\undefined}
}%
\providecommand \@ifnum [1]{%
 \ifnum #1\expandafter \@firstoftwo
 \else \expandafter \@secondoftwo
 \fi
}%
\providecommand \@ifx [1]{%
 \ifx #1\expandafter \@firstoftwo
 \else \expandafter \@secondoftwo
 \fi
}%
\providecommand \natexlab [1]{#1}%
\providecommand \enquote  [1]{``#1''}%
\providecommand \bibnamefont  [1]{#1}%
\providecommand \bibfnamefont [1]{#1}%
\providecommand \citenamefont [1]{#1}%
\providecommand \href@noop [0]{\@secondoftwo}%
\providecommand \href [0]{\begingroup \@sanitize@url \@href}%
\providecommand \@href[1]{\@@startlink{#1}\@@href}%
\providecommand \@@href[1]{\endgroup#1\@@endlink}%
\providecommand \@sanitize@url [0]{\catcode `\\12\catcode `\$12\catcode
  `\&12\catcode `\#12\catcode `\^12\catcode `\_12\catcode `\%12\relax}%
\providecommand \@@startlink[1]{}%
\providecommand \@@endlink[0]{}%
\providecommand \url  [0]{\begingroup\@sanitize@url \@url }%
\providecommand \@url [1]{\endgroup\@href {#1}{\urlprefix }}%
\providecommand \urlprefix  [0]{URL }%
\providecommand \Eprint [0]{\href }%
\providecommand \doibase [0]{https://doi.org/}%
\providecommand \selectlanguage [0]{\@gobble}%
\providecommand \bibinfo  [0]{\@secondoftwo}%
\providecommand \bibfield  [0]{\@secondoftwo}%
\providecommand \translation [1]{[#1]}%
\providecommand \BibitemOpen [0]{}%
\providecommand \bibitemStop [0]{}%
\providecommand \bibitemNoStop [0]{.\EOS\space}%
\providecommand \EOS [0]{\spacefactor3000\relax}%
\providecommand \BibitemShut  [1]{\csname bibitem#1\endcsname}%
\let\auto@bib@innerbib\@empty
\bibitem [{\citenamefont {Hartmann}\ \emph {et~al.}(2006)\citenamefont
  {Hartmann}, \citenamefont {Brandão},\ and\ \citenamefont
  {Plenio}}]{hartmannStronglyInteractingPolaritons2006}%
  \BibitemOpen
  \bibfield  {author} {\bibinfo {author} {\bibfnamefont {M.~J.}\ \bibnamefont
  {Hartmann}}, \bibinfo {author} {\bibfnamefont {F.~G. S.~L.}\ \bibnamefont
  {Brandão}},\ and\ \bibinfo {author} {\bibfnamefont {M.~B.}\ \bibnamefont
  {Plenio}},\ }\bibfield  {title} {\bibinfo {title} {Strongly interacting
  polaritons in coupled arrays of cavities},\ }\href
  {https://doi.org/10.1038/nphys462} {\bibfield  {journal} {\bibinfo  {journal}
  {Nat. Phys.}\ }\textbf {\bibinfo {volume} {2}},\ \bibinfo {pages} {849}
  (\bibinfo {year} {2006})}\BibitemShut {NoStop}%
\bibitem [{\citenamefont {Bamba}\ \emph {et~al.}(2011)\citenamefont {Bamba},
  \citenamefont {Imamoğlu}, \citenamefont {Carusotto},\ and\ \citenamefont
  {Ciuti}}]{bambaOriginStrongPhoton2011}%
  \BibitemOpen
  \bibfield  {author} {\bibinfo {author} {\bibfnamefont {M.}~\bibnamefont
  {Bamba}}, \bibinfo {author} {\bibfnamefont {A.}~\bibnamefont {Imamoğlu}},
  \bibinfo {author} {\bibfnamefont {I.}~\bibnamefont {Carusotto}},\ and\
  \bibinfo {author} {\bibfnamefont {C.}~\bibnamefont {Ciuti}},\ }\bibfield
  {title} {\bibinfo {title} {Origin of strong photon antibunching in weakly
  nonlinear photonic molecules},\ }\href
  {https://doi.org/10.1103/PhysRevA.83.021802} {\bibfield  {journal} {\bibinfo
  {journal} {Phys. Rev. A}\ }\textbf {\bibinfo {volume} {83}},\ \bibinfo
  {pages} {021802} (\bibinfo {year} {2011})}\BibitemShut {NoStop}%
\bibitem [{\citenamefont {Hodaei}\ \emph {et~al.}(2014)\citenamefont {Hodaei},
  \citenamefont {Miri}, \citenamefont {Heinrich}, \citenamefont
  {Christodoulides},\ and\ \citenamefont {{Mercedeh
  Khajavikhan}}}]{doi:10.1126/science.1258480}%
  \BibitemOpen
  \bibfield  {author} {\bibinfo {author} {\bibfnamefont {H.}~\bibnamefont
  {Hodaei}}, \bibinfo {author} {\bibfnamefont {M.-A.}\ \bibnamefont {Miri}},
  \bibinfo {author} {\bibfnamefont {M.}~\bibnamefont {Heinrich}}, \bibinfo
  {author} {\bibfnamefont {D.~N.}\ \bibnamefont {Christodoulides}},\ and\
  \bibinfo {author} {\bibnamefont {{Mercedeh Khajavikhan}}},\ }\bibfield
  {title} {\bibinfo {title} {Parity-time symmetric microring lasers},\ }\href
  {https://doi.org/10.1126/science.1258480} {\bibfield  {journal} {\bibinfo
  {journal} {Science}\ }\textbf {\bibinfo {volume} {346}},\ \bibinfo {pages}
  {975} (\bibinfo {year} {2014})}\BibitemShut {NoStop}%
\bibitem [{\citenamefont {Brandstetter}\ \emph {et~al.}(2014)\citenamefont
  {Brandstetter}, \citenamefont {Liertzer}, \citenamefont {Deutsch},
  \citenamefont {Klang}, \citenamefont {Schöberl}, \citenamefont {Türeci},
  \citenamefont {Strasser}, \citenamefont {Unterrainer},\ and\ \citenamefont
  {Rotter}}]{brandstetterReversingPumpDependence2014}%
  \BibitemOpen
  \bibfield  {author} {\bibinfo {author} {\bibfnamefont {M.}~\bibnamefont
  {Brandstetter}}, \bibinfo {author} {\bibfnamefont {M.}~\bibnamefont
  {Liertzer}}, \bibinfo {author} {\bibfnamefont {C.}~\bibnamefont {Deutsch}},
  \bibinfo {author} {\bibfnamefont {P.}~\bibnamefont {Klang}}, \bibinfo
  {author} {\bibfnamefont {J.}~\bibnamefont {Schöberl}}, \bibinfo {author}
  {\bibfnamefont {H.~E.}\ \bibnamefont {Türeci}}, \bibinfo {author}
  {\bibfnamefont {G.}~\bibnamefont {Strasser}}, \bibinfo {author}
  {\bibfnamefont {K.}~\bibnamefont {Unterrainer}},\ and\ \bibinfo {author}
  {\bibfnamefont {S.}~\bibnamefont {Rotter}},\ }\bibfield  {title} {\bibinfo
  {title} {Reversing the pump dependence of a laser at an exceptional point},\
  }\href {https://doi.org/10.1038/ncomms5034} {\bibfield  {journal} {\bibinfo
  {journal} {Nat. Commun.}\ }\textbf {\bibinfo {volume} {5}},\ \bibinfo {pages}
  {4034} (\bibinfo {year} {2014})}\BibitemShut {NoStop}%
\bibitem [{\citenamefont {Bandres}\ \emph {et~al.}(2018)\citenamefont
  {Bandres}, \citenamefont {Wittek}, \citenamefont {Harari}, \citenamefont
  {Parto}, \citenamefont {Ren}, \citenamefont {Segev}, \citenamefont
  {Christodoulides},\ and\ \citenamefont {{Mercedeh
  Khajavikhan}}}]{doi:10.1126/science.aar4005}%
  \BibitemOpen
  \bibfield  {author} {\bibinfo {author} {\bibfnamefont {M.~A.}\ \bibnamefont
  {Bandres}}, \bibinfo {author} {\bibfnamefont {S.}~\bibnamefont {Wittek}},
  \bibinfo {author} {\bibfnamefont {G.}~\bibnamefont {Harari}}, \bibinfo
  {author} {\bibfnamefont {M.}~\bibnamefont {Parto}}, \bibinfo {author}
  {\bibfnamefont {J.}~\bibnamefont {Ren}}, \bibinfo {author} {\bibfnamefont
  {M.}~\bibnamefont {Segev}}, \bibinfo {author} {\bibfnamefont {D.~N.}\
  \bibnamefont {Christodoulides}},\ and\ \bibinfo {author} {\bibnamefont
  {{Mercedeh Khajavikhan}}},\ }\bibfield  {title} {\bibinfo {title}
  {Topological insulator laser: {{Experiments}}},\ }\href
  {https://doi.org/10.1126/science.aar4005} {\bibfield  {journal} {\bibinfo
  {journal} {Science}\ }\textbf {\bibinfo {volume} {359}},\ \bibinfo {pages}
  {eaar4005} (\bibinfo {year} {2018})}\BibitemShut {NoStop}%
\bibitem [{\citenamefont {McMahon}\ \emph {et~al.}(2016)\citenamefont
  {McMahon}, \citenamefont {Marandi}, \citenamefont {Haribara}, \citenamefont
  {Hamerly}, \citenamefont {Langrock}, \citenamefont {Tamate}, \citenamefont
  {Inagaki}, \citenamefont {Takesue}, \citenamefont {Utsunomiya}, \citenamefont
  {Aihara}, \citenamefont {Byer}, \citenamefont {Fejer}, \citenamefont
  {Mabuchi},\ and\ \citenamefont {{Yoshihisa
  Yamamoto}}}]{doi:10.1126/science.aah5178}%
  \BibitemOpen
  \bibfield  {author} {\bibinfo {author} {\bibfnamefont {P.~L.}\ \bibnamefont
  {McMahon}}, \bibinfo {author} {\bibfnamefont {A.}~\bibnamefont {Marandi}},
  \bibinfo {author} {\bibfnamefont {Y.}~\bibnamefont {Haribara}}, \bibinfo
  {author} {\bibfnamefont {R.}~\bibnamefont {Hamerly}}, \bibinfo {author}
  {\bibfnamefont {C.}~\bibnamefont {Langrock}}, \bibinfo {author}
  {\bibfnamefont {S.}~\bibnamefont {Tamate}}, \bibinfo {author} {\bibfnamefont
  {T.}~\bibnamefont {Inagaki}}, \bibinfo {author} {\bibfnamefont
  {H.}~\bibnamefont {Takesue}}, \bibinfo {author} {\bibfnamefont
  {S.}~\bibnamefont {Utsunomiya}}, \bibinfo {author} {\bibfnamefont
  {K.}~\bibnamefont {Aihara}}, \bibinfo {author} {\bibfnamefont {R.~L.}\
  \bibnamefont {Byer}}, \bibinfo {author} {\bibfnamefont {M.~M.}\ \bibnamefont
  {Fejer}}, \bibinfo {author} {\bibfnamefont {H.}~\bibnamefont {Mabuchi}},\
  and\ \bibinfo {author} {\bibnamefont {{Yoshihisa Yamamoto}}},\ }\bibfield
  {title} {\bibinfo {title} {A fully programmable 100-spin coherent {{Ising}}
  machine with all-to-all connections},\ }\href
  {https://doi.org/10.1126/science.aah5178} {\bibfield  {journal} {\bibinfo
  {journal} {Science}\ }\textbf {\bibinfo {volume} {354}},\ \bibinfo {pages}
  {614} (\bibinfo {year} {2016})}\BibitemShut {NoStop}%
\bibitem [{\citenamefont {Hodaei}\ \emph {et~al.}(2017)\citenamefont {Hodaei},
  \citenamefont {Hassan}, \citenamefont {Wittek}, \citenamefont
  {Garcia-Gracia}, \citenamefont {El-Ganainy}, \citenamefont
  {Christodoulides},\ and\ \citenamefont
  {Khajavikhan}}]{hodaeiEnhancedSensitivityHigherorder2017}%
  \BibitemOpen
  \bibfield  {author} {\bibinfo {author} {\bibfnamefont {H.}~\bibnamefont
  {Hodaei}}, \bibinfo {author} {\bibfnamefont {A.~U.}\ \bibnamefont {Hassan}},
  \bibinfo {author} {\bibfnamefont {S.}~\bibnamefont {Wittek}}, \bibinfo
  {author} {\bibfnamefont {H.}~\bibnamefont {Garcia-Gracia}}, \bibinfo {author}
  {\bibfnamefont {R.}~\bibnamefont {El-Ganainy}}, \bibinfo {author}
  {\bibfnamefont {D.~N.}\ \bibnamefont {Christodoulides}},\ and\ \bibinfo
  {author} {\bibfnamefont {M.}~\bibnamefont {Khajavikhan}},\ }\bibfield
  {title} {\bibinfo {title} {Enhanced sensitivity at higher-order exceptional
  points},\ }\href {https://doi.org/10.1038/nature23280} {\bibfield  {journal}
  {\bibinfo  {journal} {Nature}\ }\textbf {\bibinfo {volume} {548}},\ \bibinfo
  {pages} {187} (\bibinfo {year} {2017})}\BibitemShut {NoStop}%
\bibitem [{\citenamefont {Boriskina}\ and\ \citenamefont
  {Negro}(2010)}]{boriskinaSelfreferencedPhotonicMolecule2010}%
  \BibitemOpen
  \bibfield  {author} {\bibinfo {author} {\bibfnamefont {S.~V.}\ \bibnamefont
  {Boriskina}}\ and\ \bibinfo {author} {\bibfnamefont {L.~D.}\ \bibnamefont
  {Negro}},\ }\bibfield  {title} {\bibinfo {title} {Self-referenced photonic
  molecule bio(chemical) sensor},\ }\href
  {https://doi.org/10.1364/OL.35.002496} {\bibfield  {journal} {\bibinfo
  {journal} {Opt. Lett.}\ }\textbf {\bibinfo {volume} {35}},\ \bibinfo {pages}
  {2496} (\bibinfo {year} {2010})}\BibitemShut {NoStop}%
\bibitem [{\citenamefont {Zhang}\ \emph {et~al.}(2015)\citenamefont {Zhang},
  \citenamefont {Liu}, \citenamefont {Wang}, \citenamefont {Gu}, \citenamefont
  {Li}, \citenamefont {Yi}, \citenamefont {Xiao},\ and\ \citenamefont
  {Song}}]{zhangSingleNanoparticleDetection2015}%
  \BibitemOpen
  \bibfield  {author} {\bibinfo {author} {\bibfnamefont {N.}~\bibnamefont
  {Zhang}}, \bibinfo {author} {\bibfnamefont {S.}~\bibnamefont {Liu}}, \bibinfo
  {author} {\bibfnamefont {K.}~\bibnamefont {Wang}}, \bibinfo {author}
  {\bibfnamefont {Z.}~\bibnamefont {Gu}}, \bibinfo {author} {\bibfnamefont
  {M.}~\bibnamefont {Li}}, \bibinfo {author} {\bibfnamefont {N.}~\bibnamefont
  {Yi}}, \bibinfo {author} {\bibfnamefont {S.}~\bibnamefont {Xiao}},\ and\
  \bibinfo {author} {\bibfnamefont {Q.}~\bibnamefont {Song}},\ }\bibfield
  {title} {\bibinfo {title} {Single {{Nanoparticle Detection Using Far-field
  Emission}} of {{Photonic Molecule}} around the {{Exceptional Point}}},\
  }\href {https://doi.org/10.1038/srep11912} {\bibfield  {journal} {\bibinfo
  {journal} {Sci. Rep.}\ }\textbf {\bibinfo {volume} {5}},\ \bibinfo {pages}
  {11912} (\bibinfo {year} {2015})}\BibitemShut {NoStop}%
\bibitem [{\citenamefont {Börgel}\ \emph {et~al.}(2016)\citenamefont
  {Börgel}, \citenamefont {Campbell},\ and\ \citenamefont
  {Ritter}}]{borgelTransitionMetalDOrbital2016}%
  \BibitemOpen
  \bibfield  {author} {\bibinfo {author} {\bibfnamefont {J.}~\bibnamefont
  {Börgel}}, \bibinfo {author} {\bibfnamefont {M.~G.}\ \bibnamefont
  {Campbell}},\ and\ \bibinfo {author} {\bibfnamefont {T.}~\bibnamefont
  {Ritter}},\ }\bibfield  {title} {\bibinfo {title} {Transition {{Metal}}
  d-{{Orbital Splitting Diagrams}}: {{An Updated Educational Resource}} for
  {{Square Planar Transition Metal Complexes}}},\ }\href
  {https://doi.org/10.1021/acs.jchemed.5b00542} {\bibfield  {journal} {\bibinfo
   {journal} {J. Chem. Educ.}\ }\textbf {\bibinfo {volume} {93}},\ \bibinfo
  {pages} {118} (\bibinfo {year} {2016})}\BibitemShut {NoStop}%
\bibitem [{\citenamefont {Kośmider}\ \emph {et~al.}(2013)\citenamefont
  {Kośmider}, \citenamefont {González},\ and\ \citenamefont
  {Fernández-Rossier}}]{kosmiderLargeSpinSplitting2013}%
  \BibitemOpen
  \bibfield  {author} {\bibinfo {author} {\bibfnamefont {K.}~\bibnamefont
  {Kośmider}}, \bibinfo {author} {\bibfnamefont {J.~W.}\ \bibnamefont
  {González}},\ and\ \bibinfo {author} {\bibfnamefont {J.}~\bibnamefont
  {Fernández-Rossier}},\ }\bibfield  {title} {\bibinfo {title} {Large spin
  splitting in the conduction band of transition metal dichalcogenide
  monolayers},\ }\href {https://doi.org/10.1103/PhysRevB.88.245436} {\bibfield
  {journal} {\bibinfo  {journal} {Phys. Rev. B}\ }\textbf {\bibinfo {volume}
  {88}},\ \bibinfo {pages} {245436} (\bibinfo {year} {2013})}\BibitemShut
  {NoStop}%
\bibitem [{\citenamefont {Zhu}\ \emph {et~al.}(2011)\citenamefont {Zhu},
  \citenamefont {Cheng},\ and\ \citenamefont
  {Schwingenschlögl}}]{zhuGiantSpinorbitinducedSpin2011a}%
  \BibitemOpen
  \bibfield  {author} {\bibinfo {author} {\bibfnamefont {Z.~Y.}\ \bibnamefont
  {Zhu}}, \bibinfo {author} {\bibfnamefont {Y.~C.}\ \bibnamefont {Cheng}},\
  and\ \bibinfo {author} {\bibfnamefont {U.}~\bibnamefont
  {Schwingenschlögl}},\ }\bibfield  {title} {\bibinfo {title} {Giant
  spin-orbit-induced spin splitting in two-dimensional transition-metal
  dichalcogenide semiconductors},\ }\href
  {https://doi.org/10.1103/PhysRevB.84.153402} {\bibfield  {journal} {\bibinfo
  {journal} {Phys. Rev. B}\ }\textbf {\bibinfo {volume} {84}},\ \bibinfo
  {pages} {153402} (\bibinfo {year} {2011})}\BibitemShut {NoStop}%
\bibitem [{\citenamefont {Fasching}\ \emph {et~al.}(2009)\citenamefont
  {Fasching}, \citenamefont {Deutsch}, \citenamefont {Benz}, \citenamefont
  {Andrews}, \citenamefont {Klang}, \citenamefont {Zobl}, \citenamefont
  {Schrenk}, \citenamefont {Strasser}, \citenamefont {Ragulis}, \citenamefont
  {Tamošiūnas},\ and\ \citenamefont
  {Unterrainer}}]{faschingElectricallyControllablePhotonic2009a}%
  \BibitemOpen
  \bibfield  {author} {\bibinfo {author} {\bibfnamefont {G.}~\bibnamefont
  {Fasching}}, \bibinfo {author} {\bibfnamefont {C.}~\bibnamefont {Deutsch}},
  \bibinfo {author} {\bibfnamefont {A.}~\bibnamefont {Benz}}, \bibinfo {author}
  {\bibfnamefont {A.~M.}\ \bibnamefont {Andrews}}, \bibinfo {author}
  {\bibfnamefont {P.}~\bibnamefont {Klang}}, \bibinfo {author} {\bibfnamefont
  {R.}~\bibnamefont {Zobl}}, \bibinfo {author} {\bibfnamefont {W.}~\bibnamefont
  {Schrenk}}, \bibinfo {author} {\bibfnamefont {G.}~\bibnamefont {Strasser}},
  \bibinfo {author} {\bibfnamefont {P.}~\bibnamefont {Ragulis}}, \bibinfo
  {author} {\bibfnamefont {V.}~\bibnamefont {Tamošiūnas}},\ and\ \bibinfo
  {author} {\bibfnamefont {K.}~\bibnamefont {Unterrainer}},\ }\bibfield
  {title} {\bibinfo {title} {Electrically controllable photonic molecule
  laser},\ }\href {https://doi.org/10.1364/OE.17.020321} {\bibfield  {journal}
  {\bibinfo  {journal} {Opt. Express}\ }\textbf {\bibinfo {volume} {17}},\
  \bibinfo {pages} {20321} (\bibinfo {year} {2009})}\BibitemShut {NoStop}%
\bibitem [{\citenamefont {Hugi}\ \emph {et~al.}(2012)\citenamefont {Hugi},
  \citenamefont {Villares}, \citenamefont {Blaser}, \citenamefont {Liu},\ and\
  \citenamefont {Faist}}]{hugiMidinfraredFrequencyComb2012a}%
  \BibitemOpen
  \bibfield  {author} {\bibinfo {author} {\bibfnamefont {A.}~\bibnamefont
  {Hugi}}, \bibinfo {author} {\bibfnamefont {G.}~\bibnamefont {Villares}},
  \bibinfo {author} {\bibfnamefont {S.}~\bibnamefont {Blaser}}, \bibinfo
  {author} {\bibfnamefont {H.~C.}\ \bibnamefont {Liu}},\ and\ \bibinfo {author}
  {\bibfnamefont {J.}~\bibnamefont {Faist}},\ }\bibfield  {title} {\bibinfo
  {title} {Mid-infrared frequency comb based on a quantum cascade laser},\
  }\href {https://doi.org/10.1038/nature11620} {\bibfield  {journal} {\bibinfo
  {journal} {Nature}\ }\textbf {\bibinfo {volume} {492}},\ \bibinfo {pages}
  {229} (\bibinfo {year} {2012})}\BibitemShut {NoStop}%
\bibitem [{\citenamefont {Kacmoli}\ \emph {et~al.}(2022)\citenamefont
  {Kacmoli}, \citenamefont {Sivco},\ and\ \citenamefont
  {Gmachl}}]{kacmoliUnidirectionalModeSelection2022}%
  \BibitemOpen
  \bibfield  {author} {\bibinfo {author} {\bibfnamefont {S.}~\bibnamefont
  {Kacmoli}}, \bibinfo {author} {\bibfnamefont {D.~L.}\ \bibnamefont {Sivco}},\
  and\ \bibinfo {author} {\bibfnamefont {C.~F.}\ \bibnamefont {Gmachl}},\
  }\bibfield  {title} {\bibinfo {title} {Unidirectional mode selection in
  bistable quantum cascade ring lasers},\ }\href
  {https://doi.org/10.1364/OE.465125} {\bibfield  {journal} {\bibinfo
  {journal} {Opt. Express}\ }\textbf {\bibinfo {volume} {30}},\ \bibinfo
  {pages} {47475} (\bibinfo {year} {2022})}\BibitemShut {NoStop}%
\bibitem [{\citenamefont {Piccardo}\ \emph {et~al.}(2020)\citenamefont
  {Piccardo}, \citenamefont {Schwarz}, \citenamefont {Kazakov}, \citenamefont
  {Beiser}, \citenamefont {Opačak}, \citenamefont {Wang}, \citenamefont {Jha},
  \citenamefont {Hillbrand}, \citenamefont {Tamagnone}, \citenamefont {Chen},
  \citenamefont {Zhu}, \citenamefont {Columbo}, \citenamefont {Belyanin},\ and\
  \citenamefont {Capasso}}]{piccardoFrequencyCombsInduced2020a}%
  \BibitemOpen
  \bibfield  {author} {\bibinfo {author} {\bibfnamefont {M.}~\bibnamefont
  {Piccardo}}, \bibinfo {author} {\bibfnamefont {B.}~\bibnamefont {Schwarz}},
  \bibinfo {author} {\bibfnamefont {D.}~\bibnamefont {Kazakov}}, \bibinfo
  {author} {\bibfnamefont {M.}~\bibnamefont {Beiser}}, \bibinfo {author}
  {\bibfnamefont {N.}~\bibnamefont {Opačak}}, \bibinfo {author} {\bibfnamefont
  {Y.}~\bibnamefont {Wang}}, \bibinfo {author} {\bibfnamefont {S.}~\bibnamefont
  {Jha}}, \bibinfo {author} {\bibfnamefont {J.}~\bibnamefont {Hillbrand}},
  \bibinfo {author} {\bibfnamefont {M.}~\bibnamefont {Tamagnone}}, \bibinfo
  {author} {\bibfnamefont {W.~T.}\ \bibnamefont {Chen}}, \bibinfo {author}
  {\bibfnamefont {A.~Y.}\ \bibnamefont {Zhu}}, \bibinfo {author} {\bibfnamefont
  {L.~L.}\ \bibnamefont {Columbo}}, \bibinfo {author} {\bibfnamefont
  {A.}~\bibnamefont {Belyanin}},\ and\ \bibinfo {author} {\bibfnamefont
  {F.}~\bibnamefont {Capasso}},\ }\bibfield  {title} {\bibinfo {title}
  {Frequency combs induced by phase turbulence},\ }\href
  {https://doi.org/10.1038/s41586-020-2386-6} {\bibfield  {journal} {\bibinfo
  {journal} {Nature}\ }\textbf {\bibinfo {volume} {582}},\ \bibinfo {pages}
  {360} (\bibinfo {year} {2020})}\BibitemShut {NoStop}%
\bibitem [{\citenamefont {Meng}\ \emph {et~al.}(2020)\citenamefont {Meng},
  \citenamefont {Singleton}, \citenamefont {Shahmohammadi}, \citenamefont
  {Kapsalidis}, \citenamefont {Wang}, \citenamefont {Beck},\ and\ \citenamefont
  {Faist}}]{mengMidinfraredFrequencyComb2020}%
  \BibitemOpen
  \bibfield  {author} {\bibinfo {author} {\bibfnamefont {B.}~\bibnamefont
  {Meng}}, \bibinfo {author} {\bibfnamefont {M.}~\bibnamefont {Singleton}},
  \bibinfo {author} {\bibfnamefont {M.}~\bibnamefont {Shahmohammadi}}, \bibinfo
  {author} {\bibfnamefont {F.}~\bibnamefont {Kapsalidis}}, \bibinfo {author}
  {\bibfnamefont {R.}~\bibnamefont {Wang}}, \bibinfo {author} {\bibfnamefont
  {M.}~\bibnamefont {Beck}},\ and\ \bibinfo {author} {\bibfnamefont
  {J.}~\bibnamefont {Faist}},\ }\bibfield  {title} {\bibinfo {title}
  {Mid-infrared frequency comb from a ring quantum cascade laser},\ }\href
  {https://doi.org/10.1364/OPTICA.377755} {\bibfield  {journal} {\bibinfo
  {journal} {Optica}\ }\textbf {\bibinfo {volume} {7}},\ \bibinfo {pages} {162}
  (\bibinfo {year} {2020})}\BibitemShut {NoStop}%
\bibitem [{\citenamefont {Meng}\ \emph {et~al.}(2022)\citenamefont {Meng},
  \citenamefont {Singleton}, \citenamefont {Hillbrand}, \citenamefont
  {Franckié}, \citenamefont {Beck},\ and\ \citenamefont
  {Faist}}]{mengDissipativeKerrSolitons2022}%
  \BibitemOpen
  \bibfield  {author} {\bibinfo {author} {\bibfnamefont {B.}~\bibnamefont
  {Meng}}, \bibinfo {author} {\bibfnamefont {M.}~\bibnamefont {Singleton}},
  \bibinfo {author} {\bibfnamefont {J.}~\bibnamefont {Hillbrand}}, \bibinfo
  {author} {\bibfnamefont {M.}~\bibnamefont {Franckié}}, \bibinfo {author}
  {\bibfnamefont {M.}~\bibnamefont {Beck}},\ and\ \bibinfo {author}
  {\bibfnamefont {J.}~\bibnamefont {Faist}},\ }\bibfield  {title} {\bibinfo
  {title} {Dissipative {{Kerr}} solitons in semiconductor ring lasers},\ }\href
  {https://doi.org/10.1038/s41566-021-00927-3} {\bibfield  {journal} {\bibinfo
  {journal} {Nat. Photon.}\ }\textbf {\bibinfo {volume} {16}},\ \bibinfo
  {pages} {142} (\bibinfo {year} {2022})}\BibitemShut {NoStop}%
\bibitem [{\citenamefont {Kim}\ \emph {et~al.}(2016)\citenamefont {Kim},
  \citenamefont {Hwang}, \citenamefont {Kim}, \citenamefont {Choi},
  \citenamefont {No},\ and\ \citenamefont
  {Park}}]{kimDirectObservationExceptional2016}%
  \BibitemOpen
  \bibfield  {author} {\bibinfo {author} {\bibfnamefont {K.-H.}\ \bibnamefont
  {Kim}}, \bibinfo {author} {\bibfnamefont {M.-S.}\ \bibnamefont {Hwang}},
  \bibinfo {author} {\bibfnamefont {H.-R.}\ \bibnamefont {Kim}}, \bibinfo
  {author} {\bibfnamefont {J.-H.}\ \bibnamefont {Choi}}, \bibinfo {author}
  {\bibfnamefont {Y.-S.}\ \bibnamefont {No}},\ and\ \bibinfo {author}
  {\bibfnamefont {H.-G.}\ \bibnamefont {Park}},\ }\bibfield  {title} {\bibinfo
  {title} {Direct observation of exceptional points in coupled photonic-crystal
  lasers with asymmetric optical gains},\ }\href
  {https://doi.org/10.1038/ncomms13893} {\bibfield  {journal} {\bibinfo
  {journal} {Nat. Commun.}\ }\textbf {\bibinfo {volume} {7}},\ \bibinfo {pages}
  {13893} (\bibinfo {year} {2016})}\BibitemShut {NoStop}%
\end{thebibliography}
\end{document}